\documentstyle[12pt,titlepage]{article}
\begin{document}

\setlength{\baselineskip}{15pt}

\begin{center}
{\Large QUANTUM--GRAVITY PATH--INTEGRALS \\ ON SIMPLICIAL LATTICES
\footnote{To be published in the Proceedings of the International
Symposium on "Quantum Physics and the Universe",
Tokyo, Japan, 19-23 Aug. 1992}} \\[1.5cm]
{\large Wolfgang Beirl, Erwin Gerstenmayer, Harald Markum} \\[1cm]
Institut f\"ur Kernphysik, Technische Universit\"at Wien, A-1040
Vienna, Austria \\[1.5cm]
\end{center}

General relativity relates classical gravitation to the curvature
of space-time. Consequently every description of quantum gravity
has to be a quantum theory of space-time geometry.
On the search for such a description the sum-over-histories
formulation leads to the Euclidean path integral
\begin{equation}
Z = \int D{\bf g}e^{-I_E({\bf g})},
\end{equation}
where the functional integration extends over a certain class of
4-geometries ${\bf g}$ with gravitational action $I_E({\bf g})$
(Hawking 1979).

Additional prescriptions are necessary to define the path integral
as a limit of systematic approximations. A direct route to such
approximations is the Regge calculus that uses simplicial lattices
to get a coordinate independent and very elegant discretization
of general relativity. It has been studied both for the classical
and quantum case and it is the starting point for our numerical
computations (Cheeger {\it et al.} 1984, Hartle 1985).

In the case of pure gravity the action is unbounded due to rapid
conformal fluctuations. This divergence is present also in the
Regge action if one does not restrict or even fix the link lengths.
However, the unboundedness of the action does not rule out a
well-defined path integral
$ Z = \int dI_E n(I_E) e^{-I_E} $
if the state density $n(I_E)$ vanishes rapidly enough for
$I_E \rightarrow -\infty$.
Indeed, previous computations indicate that the entropy of the
system could compensate the unbounded action leading to
a well-defined phase with small and negative curvature separated
from a region of unbounded positive average curvature (Berg 1986,
Hamber 1992). In the following we investigate the existence of
this phase on a regular and certain irregular triangulations of
the four-torus.

In four dimensions a simplicial lattice is a collection of $N_4$
4-simplices which are glued together to form a piecewise flat
manifold. A 4-simplex consists of 5 vertices, 10 links, 10
triangles, 5 tetrahedras and the whole lattice contains $N_d$
d-simplices. The values of $N_d$ depend on the construction of
the lattice which can be described by incidence matrices, $i.e.$
labeling all the d-simplices and specifying if and how they are
contained in each other.

Dividing a 4-torus into $N_0$ hypercubes and then each hypercube
into 24 4-simplices leads to a regular triangulation with
$N_1 = 15 N_0$, $N_2 = 50 N_0$, $N_3 = 60 N_0$ and $N_4 = 24 N_0$
(Berg 1986). The regularity of the lattice is lost if one performs
a finite number of barycentric subdivision (b.s.d.) steps afterwards.
At every b.s.d. step one chooses a 4-simplex s at random and
implants an additional vertex in its center raising the numbers
$N_d$ by an amount $\Delta N_d$ with $\Delta N_0 = 1 $,
$\Delta N_1 = 5 $, $\Delta N_2 = 10 $, $\Delta N_3 = 10 $
and $\Delta N_4 = 4 $.

After specifying the construction of the lattice one can assign
a length to each link such that all Euclidean triangle inequalities in four
dimensions are fulfilled.
Given a Euclidean configuration $\{q_l\}$ where $q_l$ is the
squared length of link $l$ one
can calculate the volume $V_s$ of each 4-simplex $s$, the area $A_t$
of each triangle $t$ and the deficit angle
\begin{equation}
\delta_t=2\pi - \sum_{s\supset t}\Theta_{s,t}.
\end{equation}
The sum runs over all 4-simplices $s$ sharing the triangle $t$
and $\Theta_{s,t}$ denotes the dihedral angle between the two tetrahedras
in simplex $s$ which have triangle $t$ in common (Hartle 1985).
Further, the fatness $\phi_s$ of each 4-simplex $s$ can be defined as
\begin{equation}
\phi_s = C^2 \frac{{V_s}^2}{max_{l\subset s}(q_l^4)},
\end{equation}
where we set the constant $C = 24$.

The Einstein-Hilbert action can be replaced then by a sum over all triangles
in the simplicial lattice
\begin{eqnarray}
 L_P^{-2} \int d^4x g^{\frac{1}{2}}R \longrightarrow
 L^{-2}_P \sum_t 2A_t \delta_t ,
\end{eqnarray}
where $L_P$ is the Planck length, $R$ the curvature scalar and $g$ the
determinant of the metric.
One can show that the classical continuum limit can be reached if the
fatness of each 4-simplex stays larger than zero, $i.e.$
$\phi_s > f = const > 0 $ (Cheeger {\it et al.} 1984).

In our investigation of quantum gravity we consider only
four-geometries with finite total volume and use therefore the Regge action
with a cosmological constant term
\begin{equation}
-I_E = \beta \sum_t A_t \delta_t - \lambda \sum_s V_s .
\end{equation}
We measure all geometric quantities in units of $L_P$ and rescale
the lattice such that $\lambda = 1$.

One can now use the path integral of simplicial quantum gravity
to calculate physical expectation values (Hartle 1985).
Since we hold the incidence matrices of the lattice fixed and vary only
the $q_l$ the integration (1) reduces to a summation over different
configurations $\{q_l\}$.
As in the continuum case a unique definition of the measure in the
path-integral is not known.
Previous simulations indicate that a uniform measure
$ D \mu = ( \prod_l {dq_l}) {\cal F} (q_1,...,q_{N_1}) $
could lead to reasonable results and we use it in the
following (Beirl {\it et al.} 1992). The function $\cal{F}$ is one for
Euclidean
configurations $\{q_l\}$ and zero otherwise.

As in the continuum the Regge action $I_E$ is not bounded because
the sum $\sum_t A_t \delta_t$
suffers from a divergence that occurs when some 4-simplices of the
lattice collapse. Although the
4-volume $\sum_s V_s$ of the lattice remains finite this leads to
triangles with
$\delta_t > 0$ and $A_t \rightarrow \infty$ and thus the Regge
action runs to minus infinity.
Such a collapse of simplices was observed in previous simulations as
a consequence of a formation of spikes and seems to reflect rapid and
large conformal fluctuations (Beirl {\it et al.} 1992).

\begin{figure}[t]
\input{fig1}
{ \setlength{\baselineskip}{12pt} \small
FIG. 1. Average curvature $R$ as a function of $\beta$
on a regularly triangulated 4-torus with $4^4$ vertices for
decreasing cutoff $f$ (a) and for
$B = 0, 5, 10$ and 20 irregularly inserted vertices (b). Averages
are taken over at least 2500 iterations for each point after thermalization.
Error bars due to mean standard deviation are smaller than the symbols.
}
\end{figure}

\begin{figure}[t]
\input{fig2}
{ \setlength{\baselineskip}{12pt} \small
FIG. 2. Distances $d_v$ of the vertices from their next neighbours.
The upper two pictures (a) show configurations of the regularly triangulated
4-torus below and above $\beta_0$. The lower plots (b) reflect
configurations of the irregularly triangulated 4-torus with 5
b.s.d. steps; the corresponding vertices are inserted into the figure
at equidistant separations for better visibility.
}
\end{figure}

However, the fatness $\phi_s$ of a collapsing 4-simplex tends to zero
and this was the motivation to restrict the domain of integration by
incorporating the requirement $\phi_s > f > 0 $ into the measure.
The function  ${\cal F}_f (q_1,...,q_{N_1})$ is now zero for all
configurations violating either the Euclidean triangle inequalities or
the lower limit on the fatness. With this restriction the action is bounded
for all $f > 0$ as long as the number of 4-simplices $N_4$ is finite.

To study the properties of this system for different values of $f$
we carried out numerical calculations on a regularly triangulated 4-torus
with $4^4$ vertices.
The Monte Carlo computations proceed in the usual way as described in
the literature (Hamber 1992). We performed up to 20000 Metropolis
sweeps for each data point and took the averages after thermalization.

In figure 1a the average curvature of the lattice
$ R = \langle q_l \rangle \frac{ \langle \sum_t A_t \delta_t \rangle}
 { \langle \sum_s V_s \rangle } $
is plotted versus the coupling parameter $\beta$  for the constant $f$
decreasing from $10^{-3}$ down to $10^{-5}$. For $0 \leq \beta < \beta_0(f)$
the average curvature $R$ stays small and negative. Although the
gravitational action favours configurations with large positive
curvature they occur for $\beta > \beta_0$ only.
It is important that we observe no hysteresis and the computed
expectation values seem to be independent of the start configuration.
For the smallest value $f=10^{-5}$
our results agree with calculations for the pure Regge action
by Hamber (1992).

The configurations change in a characteristic way
across the transition at $\beta_0$. We computed for some configurations
the next neighbour distances $d_v$ by averaging ${q_l}^\frac{1}{2}$ over
all links meeting at the same vertex $v$. Figure 2a shows these distances
below and above $\beta_0$ for $f = 5{\times}10^{-4}$.

To study the dependence of the phase structure on the underlying
triangulation we applied a (small) number $B$ of
b.s.d. steps on the regular lattice with $4^4$ vertices creating
irregularities as described above.
Applying again a lower limit $f$ on the fatness numerical computations are
possible and show a shift to positive curvature with increasing $B$.
Figure 1b presents the results for $B = 0, 5, 10$ and 20 with
$f = 5{\times}10^{-4}$. The existence of two phases indicates a
qualitatively similar
behaviour of the regularly and irregularly triangulated lattices.
The shift of the curvature originates from positive contributions of the
inserted vertices.
This is seen directly in the configurations. The additional vertices tend to
develope a spike as displayed in figure 2b showing the next neighbour
distances $d_v$ for $B = 5$ and $f = 5{\times}10^{-4}$ while the other
vertices behave like those of the regular lattice.

To finish, we investigated quantum gravity within the Regge calculus both for
regularly and irregularly triangulated lattices. The first main question under
consideration was the existence of a well-defined phase. Our results
exhibit such a phase with small and negative mean curvature for both types
of triangulations. The calculations on irregular lattices may be regarded
as a first step towards computations on a random lattice.
While for restricted fatness the existence of a well-defined phase seems to
be established we did not adress the limit $f \rightarrow 0$.
In this case configurations with large spikes could play an
important role but may be reached only for extreme statistics.
Therefore, the question whether the entropy of the system can even compensate
the unbounded pure Regge action is difficult to answer and subject of
current investigations.

This work was supported in part by "Fonds zur F\"orderung der
wissenschaftlichen Forschung" under Contract No. P7510-TEC.

\setlength{\baselineskip}{12pt}
\begin{center}
REFERENCES
\end{center}

\noindent
{\small
Beirl, W., Gerstenmayer, E., Markum, H. (1992) Influence of the Measure
on Simplicial Quantum Gravity in Four Dimensions. { \it Phys. Rev.
Lett.} {\bf 69}, 713-716. \\[0.1cm]
Berg, B. (1986) Entropy versus Energy on a Fluctuating Four-Dimensional
Regge Skeleton. { \it Phys. Lett. B} {\bf 176}, 39-44. \\[0.1cm]
Cheeger, J., M\"uller, W., Schrader, R. (1984) On the Curvature of
Piecewise Flat Spaces. { \it Comm. Math. Phys.} {\bf 92}, 405-454. \\[0.1cm]
Hamber, H. (1992) Phases of Four-Dimensional Simplicial Quantum Gravity.
{ \it Phys. Rev.  D} { \bf 45}, 507-512. \\[0.1cm]
Hartle, J. (1985) Simplicial Minisuperspace. I.~General Discussion. { \it
J. Math. Phys.} {\bf 26}, 804-814. \\[0.1cm]
Hawking, S. (1979) The Path-Integral Approach to Quantum Gravity. In
{\it General Relativity - An Einstein Centenary Survey}, Eds. Hawking,
S. and Israel, W., pp. 746-789.  Cambridge University Press, Cambridge. \\
}

\end{document}